\documentstyle[12pt,axodraw]{article}
\def\beq{\begin{equation}}
\def\eeq#1{\label{#1}\end{equation}}
\def\eeqn{\end{equation}}
\def\beqa{\begin{eqnarray}}
\def\eeqa#1{\label{#1}\end{eqnarray}}
\def\eeqan{\end{eqnarray}}
\def\CR{\nonumber \\ }
\def\leqn#1{(\ref{#1})}

\newcommand{\lsim}{\, \mathop{}_{\textstyle \sim}^{\textstyle <} \,}

\def\goesto{\rightarrow}
\def\4pl{{\bf 4}}
\def\2pl{{\bf 2}}

\begin{document}
\baselineskip 0.6cm

\begin{titlepage}

\begin{flushright}
UCB-PTH-02/46 \\
LBNL-51613 \\
hep-ph/0210149 \\
\end{flushright}

\vskip 1.0cm

\begin{center}
{\Large \bf The Weak Mixing Angle From TeV Scale Quark-Lepton Unification}

\vskip 1.0cm

{\large Z. Chacko,
Lawrence J.~Hall, 
and Maxim Perelstein}

\vskip 0.5cm

{\it Department of Physics, University of California,
                Berkeley, CA 94720, USA}\\
{\it Theoretical Physics Group, Lawrence Berkeley National Laboratory,
                Berkeley, CA 94720, USA}

\vskip 1.0cm

\abstract{
Unified theories based on an extended left-right symmetric group, 
$SU(4) \times SU(2)^4$, are constructed in five dimensions. The 
compactification scale is assumed to be only a loop factor above the weak 
scale, so that the weak mixing angle is predicted to be close to its tree 
level value of 0.239. Boundary conditions in the 5th dimension break
\if
Unification of quarks and leptons via left-right symmetry, 
$SU(4) \times SU(2)_L \times SU(2)_R$, is extended to $SU(4) \times 
SU(2)^4$. All four $SU(2)$ interactions have the same gauge coupling,
leading to a tree-level prediction of the weak mixing angle, $\sin^2
\theta = 0.239$, which is so close to data that this unification must 
occur around the TeV scale. 
Five dimensional theories are constructed with boundary condition
breaking of 
\fi
$SU(4) \rightarrow SU(3) \times U(1)_{B-L}$, removing
powerful constraints from $K_L \rightarrow \mu e$
while allowing a reliable calculation of the leading logarithm
corrections to $\sin^2 \theta$. 
The compactification scale is expected in the 1--5 TeV region,
depending on how $SU(2)^4$ is broken. Two illustrative models are
presented, and the experimental signal of the $Z'$ gauge boson is
discussed.}

\end{center}
\end{titlepage}

\section{Introduction} \label{sec:intro}

The strong, weak and electromagnetic forces, so differently manifested
in nature, are described by
underlying gauge interactions based on common principles. From this
similarity in structure it becomes highly plausible that these three
interactions are just low energy remnants of a more unified gauge
theory. Such a unified theory could provide predictions for ratios of
gauge couplings and ratios of fermion masses, as well as an understanding
of the quark and lepton quantum numbers.

The most basic question about such a unification of the gauge forces
is the mass scale at which the unification occurs. The standard
paradigm of unification into a supersymmetric $SU(5)$ or $SO(10)$
theory~\cite{susyguts} has a high unification scale, of order $10^{16}$ 
GeV, and leads to the highly successful prediction for the weak mixing 
angle of $\sin^2 \theta = 0.233 \pm 0.002$. Objections to the simplest 
model, about proton decay, the lightness of the Higgs doublets compared to 
their color triplet partners and quark-lepton mass relations, can all be 
overcome by promoting $SU(5)$ to a five-dimensional (5D) gauge symmetry 
broken by boundary 
conditions in a 5th dimension, which simultaneously improves the prediction 
of the weak mixing angle to $\sin^2 \theta = 0.2313 \pm 0.0004$~\cite{HN5}. 
Although the unification mass scale is reduced to near $10^{15}$ GeV, it is
still extremely large.

It has recently been argued that a successful prediction of the weak
mixing angle is also possible if the unification occurs at a low scale, 
in the TeV domain~\cite{DK}. Even though this unification only involves
two of the Standard Model gauge couplings, the requirement that the 
unification scale be related to the electroweak symmetry breaking scale
by a loop factor makes the model predictive.
It is well known that the $SU(2) \times U(1)$ 
electroweak  theory can be embedded into $SU(3)_{EW}$ in a straightforward 
way for leptons~\cite{W}. In this scheme, hypercharge is identified as the 
diagonal generator of $SU(3)$ which is orthogonal to $SU(2)_L$:
\begin{equation}
\frac{Y}{2} = \sqrt{3} \, T_8,
\label{eq:ysu3}
\end{equation}
leading to the tree level prediction for the weak mixing angle, 
$\sin^2 \theta = 0.25.$ This prediction is intriguingly close to the
experimental value of $\sin^2 \theta$ at the $Z$ pole~\cite{pdg},
\beq
\sin^2 \theta (M_Z)|_{exp} = 0.23113 \pm 0.00015.
\eeq{eq:atmz} 
Refining the tree-level prediction of the $SU(3)_{EW}$ model by  
including one-loop radiative corrections leads to a correct prediction 
of $\sin^2 \theta (M_Z)$ provided that the unification scale (the scale
at which $SU(3)_{EW}$ breaks down to $SU(2)\times U(1)$) is about 4 TeV.
The main difficulty of this unification scheme is that the quarks do not 
exhibit any $SU(3)$ pattern. This problem can be overcome if $SU(2)_L 
\times U(1)_Y$ is embedded into $SU(2) \times U(1) \times SU(3)$ in such a 
way that the $SU(3)$ factor is most important in determining the value of
the low energy couplings~\cite{DK}. Alternatively one can construct an
$SU(3)_{EW}$ theory in 5D, with boundary condition breaking to $SU(2)_L
\times U(1)_Y$~\cite{others1, HN3, others2}. The quarks live on the boundary 
where only the $SU(2)_L \times U(1)_Y$ gauge symmetry is operative, while 
the leptons and Higgs may feel the full $SU(3)_{EW}$ gauge 
symmetry\footnote{For subsequent work along the lines of $SU(3)_{EW}$ 
unification see, for example, Refs.~\cite{DK3, Frampton}.}.

The $SU(3)_{EW}$ model of TeV scale unification can be criticized
on the grounds that the prediction for the weak mixing angle is less precise 
than in the case of high scale unification. Also, this unification
scheme gives no understanding of the quantum numbers of the 
quarks\footnote{The four-dimensional version of the model~\cite{DK} is also 
subject to severe electroweak precision constraints~\cite{pEW}.}.
Nevertheless, it provides a testable alternative to the high-scale 
unification, and becomes especially interesting if the hierarchy problem 
is resolved by lowering the fundamental gravity scale to the TeV 
region~\cite{ADD} or by introducing a warped extra dimension~\cite{RS}.

In this paper we introduce a new idea for TeV scale unification. Our
starting point is the model of Pati and Salam~\cite{PS} based on the
$SU(4) \times SU(2)_L \times SU(2)_R$ gauge group. This model provides
a very satisfactory understanding of both quark and lepton quantum numbers.
Moreover, the additional gauge bosons of this model do not lead to proton 
decay, and the constraints on the unification scale are much weaker than
for the $SU(5)$ or $SO(10)$ gauge groups. By itself, however, the 
Pati-Salam model does not provide a viable scheme for TeV scale unification. 
In this model, hypercharge is obtained as a linear combination of the 
$B-L$ generator of $SU(4)$ and the diagonal generator of $SU(2)_R$,
leading to a grossly incorrect tree-level prediction for $\sin^2 \theta$:
\begin{equation}
Y = T_{B-L} + T_{3R}  \hspace{.5in} \Longrightarrow \hspace{.5in}
 \sin^2 \theta = {1 \over 2} -
{1 \over 3} \left. {\alpha \over \alpha_s} \right|_{M_Z} = 0.478,
\label{eq:y422}
\end{equation}
where we have used the values of the Standard Model (SM) gauge couplings at 
$M_Z$\footnote{In deriving Eq.~\leqn{eq:y422}, we have treated the $SU(2)$ 
and $SU(4)$ couplings as independent, and used their experimental values
at $M_Z$ as inputs. An alternative is to assume that the Pati-Salam
group is embedded in $SO(10)$. In this case, the model predicts 
$\sin^2 \theta =3/8$ at the unification scale, which is also unacceptable 
for TeV scale unification.}. 
(We have imposed an additional $Z_2$ symmetry interchanging the $SU(2)$
factors; without such a symmetry, the model does not make a 
definite prediction for $\sin^2 \theta$.) Clearly, radiative corrections
cannot render the prediction~\leqn{eq:y422} consistent with the 
experimental result~\leqn{eq:atmz} if the unification scale is in the 
TeV range. Following the proposal of
Hung, Buras and Bjorken~\cite{petite}, we enlarge the gauge group of the 
Pati-Salam model to include two additional $SU(2)$ factors, 
which we will refer to as $SU(2)_1$ and $SU(2)_2$. The Standard 
Model quarks and leptons are not charged under the additional $SU(2)$'s; as 
we discuss below, the model may or may not contain additional matter charged 
under these groups. We assume a discrete symmetry that interchanges the four 
$SU(2)$ factors, making their (ultraviolet) gauge couplings identical. 
Crucially, the hypercharge arises as a linear combination of the $B-L$ 
generator of $SU(4)$ and the $T_3$ generator of the {\it diagonal subgroup} 
of the $SU(2)_R\times SU(2)_1 \times SU(2)_2$. This embedding of the 
hypercharge leads to the tree-level prediction for the weak mixing angle
which is remarkably close to experiment:
\begin{equation}
Y = T_{B-L} + T_{3R} + \sum_{i=1}^2 T_{3,i} \hspace{.5in} \Longrightarrow 
\hspace{.5in}  \sin^2 \theta = {1 \over 4} -
{1 \over 6} \left. {\alpha \over \alpha_s} \right|_{M_Z} = 0.239.
\label{eq:y42222}
\end{equation}

The group $SU(4)\times SU(2)^4$, with the above embedding of the Standard
Model generators, was studied in~\cite{petite} for unification at or above
the 1000 TeV scale. Breaking $SU(4)$ below this scale is generally 
excluded by limits on rare decay modes such as $K_L\goesto\mu e$~\cite{VW}.
However, since this model contains two independent fundamental gauge 
couplings, it can only be predictive if the unification scale is fixed by
independent considerations. In this work, we will require that the 
unification occur at scales between 1 and 10 TeV, about one
loop factor above the electroweak scale. This assumption is particularly
well motivated in theories with low fundamental scale and large extra
dimensions~\cite{ADD}. We will show that in the presence of extra 
dimensions the bounds from $K_L\goesto\mu e$ decays can be relaxed, leading
to a consistent and predictive theory of TeV-scale quark-lepton 
unification.    

\section{Boundary Condition Breaking of $SU(4)$} \label{bc}

There are three immediate objections to realising the idea outlined above: 
\begin{itemize}
\item As we explained at the end of Section 1, we are interested in 
theories which unify at, or slightly above, the TeV scale.
This requirement seems to contradict the experimental 
lower bounds of about 1000 TeV on the mass of the exotic $SU(4)/(SU(3) \times 
U(1)_{B-L})$ gauge bosons $X$. (A particularly strong constraint arises from 
the non-observation of $K_L \rightarrow \mu e$~\cite{VW}.)
\item A Standard Model generation consists of two $SU(4)$ 4-plets:  $\psi_L$, 
which is a doublet under $SU(2)_L$, and $\psi_R$, which is a doublet under 
$SU(2)_R$. There are no fields charged under the additional two $SU(2)$'s.
Hence, the simplest interpretation of a generation
does not allow a discrete symmetry which ensures equality of the four
$SU(2)$ gauge couplings, as needed for a prediction of the weak mixing
angle.  
\item The $SU(4)$ symmetry constrains the up quarks and neutrinos to
have Yukawa couplings of the same size. Even if the right handed
neutrinos receive Majorana masses at the TeV scale, the three light
neutrinos all have masses far in excess of experimental limits.
\end{itemize}

Although there may be several answers to these objections, in this
paper we pursue just a single idea. We assume that the symmetry
breaking  $SU(4) \rightarrow SU(3) \times U(1)_{B-L}$ is accomplished
by boundary conditions in a compact extra dimension $x_5$ of physical length
$\pi R$, with $R \sim$ TeV$^{-1}$. Explicitly, we start with a 5D $SU(4)$ 
gauge field $A_M \equiv A_M^a
T^a, M=1 \ldots 5, a=1\ldots 15$, and impose the following boundary 
conditions:
\beqa 
A_\mu (x^\mu, x_5) &=& +A_\mu (x^\mu, -x_5) = Z\,A_\mu (x^\mu, x_5+2\pi R)\,
Z^{-1}, \CR 
A_5 (x^\mu, x_5) &=& -A_5 (x^\mu, -x_5) = Z\,A_5 (x^\mu, x_5+2\pi R)\,
Z^{-1}, 
\eeqa{eq:orbifold}
where $\mu=1\ldots 4$ and $Z=$diag$(+,+,+,-)$. The low-energy effective
field theory contains nine four-dimensional (4D) massless gauge bosons 
$A_\mu$ of the 
$SU(3)\times U(1)_{B-L}$ group, while the remaining six gauge bosons $X_\mu$
and  the fifteen 4D scalars $A_5^a$ do not possess zero modes. We assume 
that the gauge bosons of the four $SU(2)$ groups of our model are free 
to propagate in the bulk; at this point,
we leave open the question as to whether $SU(2)_R \times SU(2)_1
\times SU(2)_2$ is broken by boundary conditions or by Higgs vevs. These 
assumptions have the virtue of overcoming all three objections in an 
economical way:
\begin{itemize}
\item The constraints on the $X$ boson mass are naturally avoided if the 
matter fields live in the bulk. Consider a 5D 4-plet $\Psi = (Q, L)$. Due 
to the boundary condition breaking of $SU(4)$, only one component of $\Psi$ 
has a zero mode. (This component can be either $L$ or $Q$, depending on the
charge of $\Psi$ under the reflection.) Thus, one generation of SM fermions 
requires {\it four} 5D 4-plets: $\Psi_L = (Q_L, 
\tilde{L}_L)$, $\Psi'_L = (\tilde{Q}_L, L_L)$, $\Psi_R = (Q_R, \tilde{L}_R)$, 
and $\Psi'_R = (\tilde{Q}_R, L_R)$, where the tildes mark the fields that do 
not possess zero modes. The SM quarks and leptons of the same generation 
{\it do not} come from the same $SU(4)$ multiplet, and are not coupled 
through the $X$ bosons of $SU(4)$. (This lack of unification does not destroy 
the understanding of the quantum numbers of a generation provided by $4-2-2$!)

Alternatively, SM generations can be built out of four-dimensional
fields living on a boundary of space-time where the $SU(4)$ symmetry is broken 
to $SU(3)\times U(1)$ (the ``3-1 point''). The $X$ boson wave function 
vanishes at this boundary. Note, however, that in this case the $4-2-2$ 
pattern of the quark and lepton quantum numbers is purely accidental. 
\item If quarks and leptons live in the bulk, a discrete symmetry relating 
the four $SU(2)$ factors can be imposed provided that we introduce three 
additional generations of fermions. Each additional generation consists of 
four $SU(4)$ 4-plets, two of them transforming as doublets under $SU(2)_1$ 
and the other two under $SU(2)_2$. The extra fermions acquire masses at 
the $SU(2)_1\times SU(2)_2$ breaking scale, and are therefore sufficiently 
heavy to escape detection. 

If quarks and leptons live on the 3-1 point, it is not necessary to introduce
new fields. In this case, a discrete symmetry ensuring the equality of the 
four $SU(2)$ couplings can be imposed in the bulk. This symmetry has to
be broken at the boundary, but the corrections to the weak mixing angle 
prediction due to this breaking are suppressed by the volume of the fifth 
dimension. As we explain below, this volume (in units of the fundamental
scale) is taken to be large in our model, making these corrections 
irrelevant.
\item We take the Yukawa couplings to be located on the 3-1 boundary. If the
fermions live on this boundary, this is automatic. If the fermions live in 
the bulk, the absence of bulk Yukawa couplings could be due to supersymmetry, 
or to the fact that the Higgs field is localized on the boundary. Since only 
the $SU(3) \times U(1)_{B-L}$ subgroup of the $SU(4)$ gauge interactions are 
operative at the 3-1 point, there is no relation between the Yukawa couplings 
of the up quarks and the neutrinos.
\end{itemize}

Promoting the gauge symmetry of our model to 5D raises an important issue. 
Since a 5D gauge theory is non-renormalizable, there has to be a fundamental
scale $M_s$ at which it breaks down. As we will see below, in our model 
$M_s \sim 100$ TeV. While proton stability at the renormalizable level is
ensured by the accidental symmetries of the Pati-Salam model, higher-dimension
operators induced at $M_s$ could lead to proton decay. To prevent that,
we will impose global $B$ and $L$ symmetries. Since quarks and leptons come
from different $SU(4)$ multiplets, these symmetries commute with the
gauge transformations. There is also a possibility of additional flavor
violating effects (e.g. $K_L\goesto\mu e$ decays) induced by the 
non-renormalizable operators generated at $M_s$. However, since the scale is
rather high, even a very modest amount of flavor symmetry in the fundamental
theory (or small fine tuning) would be sufficient to render these effects
unobservable at present.   

\section{Radiative Corrections and Uncertainties in the Prediction of
the Weak Mixing Angle}
\label{loops}

Does the tree-level prediction (\ref{eq:y42222}) survive in the 5D theory 
with boundary condition breaking of $SU(4)$? At tree level, the 
4D gauge couplings are given by
\beq
\frac{1}{g_{i,4}^2} = \frac{\pi R}{g_{i,5}^2} + \frac{1}{\tilde{g_i}^2},  
\eeq{eq:4dcoupling}
where $g_{i,5}$ is the corresponding 5D gauge coupling, and $1/\tilde{g_i}^2$ 
is the coefficient of the gauge kinetic term induced on the $3-1$ boundary. 
The equality of the four $SU(2)$ couplings and the equality of the strong 
coupling and the appropriately normalized $U(1)_{B-L}$ coupling, which were 
crucial in obtaining the result~\leqn{eq:y42222}, relied on the symmetries 
which are realized in the bulk but not at the $3-1$ boundary. Therefore, they 
hold for the terms involving $g_{i,5}$ but not for the boundary-induced
terms proportional to $1/\tilde{g_i}^2$. The boundary-induced terms could
therefore lead to large corrections to~\leqn{eq:y42222}. Such effects were 
also present in the 5D $SU(5)$~\cite{HN5} and the 5D $SU(3)_{EW}$~\cite{HN3} 
unification schemes, and we follow the assumptions made there to recover a 
high degree of predictivity: we take the 5D theory as the correct 
effective theory up to the scale $M_s$ at which all bulk and boundary 
gauge interactions are assumed to become strong. The strong bulk coupling 
assumption implies the existence of a hierarchy between $M_s$ and the 
compactification scale $M_c=1/R$: $M_s/M_c \approx l_5/(\pi C_2 g^2) \approx 
32\pi^2/(C_2 g^2) \gg 1$, where $C_2$ and $g$ are the quadratic Casimir and 
the low-energy gauge coupling, respectively, and $l_5=32\pi^3$ is the 5D loop 
factor~\cite{method}. The precise magnitude of the hierarchy is rather 
uncertain due to an approximate nature of the strong coupling argument. In 
our numerical estimates, we will use $M_s/M_c = 80$; this estimate is 
probably valid to about a factor of two. According to~\leqn{eq:4dcoupling}, 
this means that the boundary gauge kinetic terms are subdominant to the 
bulk ones. 

The leading logarithmic corrections to the formula~\leqn{eq:y42222} can
be readily calculated. We will make a simplifying assumption that the gauge 
symmetry breaking of $SU(2)_R \times SU(2)_1 \times SU(2)_2 \goesto T_{3R}$ 
involves no mass scale other than $M_s$ and $M_c$. With these assumptions, 
the prediction for the weak mixing angle at leading logarithm is  
\begin{eqnarray}
\sin^2 \theta(M_Z) =  {1 \over 4} 
                - {1 \over 6} \left. {\alpha \over \alpha_s} \right|_{M_Z} 
               &-& {\alpha(M_Z) \over 8 \pi}
                    \left( b_Y - 3b_2 - {2 \over 3} b_3 \right) \ln {M_c' 
\over M_Z} 
\nonumber \\
               &-& {\alpha(M_Z) \over 8 \pi}
               \left( (b_{B-L}' - {2 \over 3} b_3') 
           + (b_R' - 3 b_2') \right) 
\ln {M_s \over M_c'},
\label{eq:s^2master}
\end{eqnarray}
where $b_{3,2,Y}$ are the beta function coefficients for $g_{3,2,Y}$
below the modified compactification scale $M_c' = M_c/\pi$~\cite{HN5, HN3}, 
while $b_{3,2,B-L}'$ are the beta function coefficients for the relative
logarithmic running of the QCD, $SU(2)_L$, and $B-L$  gauge
couplings above $M_c'$. The formula~\leqn{eq:s^2master} is sufficiently
general to be used for various patterns of the $SU(2)_R \times SU(2)_1 
\times SU(2)_2 \goesto T_{3R}$ symmetry breaking: this model dependence
is encoded in the coefficient $b_R'$. For example, if the breaking occurs 
entirely by the Higgs mechanism at $M_c$, then $b_R'$ is given by the
sum of the beta function coefficients describing the relative logarithmic
running of the $SU(2)_R$, $SU(2)_1$, and $SU(2)_2$ gauge couplings
above $M_c'$; if the breaking occurs entirely by the Higgs mechanism at 
scale $M_s$, $b_R'$ is just equal to the beta function coefficient for 
the evolution of the single gauge coupling $g_{T_{3R}}$; etc.

Above the compactification scale, the gauge couplings run according to a 
power law; however, the leading corrections to the $\sin^2\theta$ prediction 
are logarithmic. This is due to {\it universality} of the power-law running: 
since the discrete symmetry relating the four $SU(2)$ factors of our model is 
broken only locally in the 5D bulk, the four gauge couplings have identical 
power law running. By the same argument, the boundary condition breaking of 
$SU(4)$ ensures that the power running of the $SU(3)$ and the (appropriately 
normalized) $U(1)_{B-L}$ coupling constants is identical, and all the power 
running effects cancel out in Eq.~\leqn{eq:s^2master}.

If the theory below $M_c$ is the Standard Model, 
the running between this scale and $M_Z$ corrects the weak mixing angle by 
$\delta \sin^2 
\theta = - 0.0065 \ln (M_c'/M_Z)$. The rules for computing the beta function 
coefficients above the compactification scale in orbifold models were given 
in~\cite{HN5}. Applying these rules to running of the $SU(3)$ and $U(1)_{B-L}$ 
gauge couplings gives $b_{B-L}' - (2/3) b_3' = 23/6$. Since matter fields come
in complete $SU(4)$ multiplets, they do not contribute to this combination  
of beta functions; therefore, this result does not depend on whether the
matter fields live in the bulk or on the boundary. It is also clearly   
independent of the pattern of the $SU(2)_R\times SU(2)_1\times SU(2)_2 
\rightarrow T_{3R}$ breaking. Inserting these results into~\leqn{eq:s^2master},
we obtain
\beq
\sin^2 \theta(M_Z) = 0.234 - 0.0065\,\ln (M_c'/M_Z) - 0.0015\,(b_R' - 3 b_2')
\pm 0.005.
\eeq{s^2num}
The $b_R' - 3 b_2'$ term in this equation is the 
model-dependent contribution from relative running of the $SU(2)$ gauge 
couplings. We will evaluate it in some explicit models in the next section. 
Before doing that, however, let us make the following important observation.
The second term of Eq.~\leqn{s^2num} is negative. If the third term is 
non-positive, $b_R' - 3 b_2' \geq 0$, reproducing the correct value of the 
weak mixing angle requires a very low value of $M_c$, inconsistent with 
experimental constraints. Thus, the successful models will be those with  
$b_R' - 3 b_2'$ negative, tending to increase $M_c$.

It is also interesting to consider the supersymmetric version of our model,
with the supersymmetry breaking scale in the visible sector 
$\tilde{m}\lsim M_c$. In this
case, we obtain $b_{B-L}' - (2/3) b_3' = 4$, and Eq.~\leqn{eq:s^2master}
becomes
\beq
\sin^2 \theta(M_Z) = 0.234 - 0.0031\,\ln (M_c'/M_Z) - 0.0015\,(b_R' - 3 b_2')
\pm 0.005,
\eeq{s^2numSUSY}
where we have not included supersymmetric threshold corrections involving 
$\ln(\tilde{m}/M_Z)$. Note that the coefficient of the second term in 
this equation is smaller than in its non-supersymmetric counterpart, 
implying that it is easier to construct realistic models in the 
supersymmetric case.

The uncertainties in the weak mixing angle predictions~\leqn{s^2num} 
and~\leqn{s^2numSUSY} come from two sources. First, as we already explained,
the predictions can be corrected at tree level by the gauge kinetic terms 
induced on the 3-1 boundary. The dominant effect comes from the violation of 
the discrete symmetry relating the four $SU(2)$ gauge couplings on that 
boundary. Using Eq.~\leqn{eq:4dcoupling} and the strong coupling assumption,
we estimate that the boundary terms correct each of the low energy gauge
couplings by an amount $\delta_i \equiv \delta g_i^2/g_i^2 \approx l_4/(l_5 
M_s \pi R) \approx 1/40$, where $l_4=16\pi^2$ is the 4D loop 
factor~\cite{method}. The uncertainty in the 
tree-level prediction for the weak mixing angle is given by $\delta\sin^2
\theta/\sin^2\theta\,\approx\,0.25\sum_i \delta_i$; assuming that the 
uncertainties from the four $SU(2)$'s add in quadruture, we arrive at an 
estimate $\delta\sin^2\theta \approx 0.003$. The second 
source of uncertainty is the non-logarithmically enhanced loop contributions 
from running between $M_s$ and $M_c'$; since $\ln(M_s/M_c') \approx 5$, we   
estimate this uncertainty to be 20\% of the term in Eq.~\leqn{eq:s^2master}
involving this logarithm. Numerically, this corresponds to $\delta\sin^2
\theta\approx 0.002$, slightly below $1\%$. Thus, we
estimate the total uncertainty in our prediction of $\sin^2\theta(M_Z)$ to be
about 2\%, or 0.005. Note that this theoretical uncertainty 
can be reduced in specific models. For example, if the discrete symmetry 
relating the four $SU(2)$ gauge couplings is only broken {\it spontaneously}
on the 3-1 boundary, it will be respected by the boundary gauge kinetic terms,
and they will not modify the weak mixing angle prediction. However, we will
not pursue this possibility in this paper.

\section{Sample Models} 
\label{models} 

In this section, we present two explicit models realizing the ideas discussed 
above which lead to acceptable predictions for the weak mixing angle. In both 
models, the $SU(4) \rightarrow SU(3) \times U(1)_{B-L}$ breaking is achieved 
by imposing boundary conditions as described in section~\ref{bc}. Both models 
are supersymmetric, with supersymmetry broken spontaneously at scales of 
order TeV. A 
5D gauge supermultiplet consists of a symplectic Majorana spinor $\lambda_i$ 
and a real scalar $\sigma$, in addition to the 5D vector field $A_M$. The 
boundary conditions for the fields in the $SU(4)$ gauge multiplet are given 
by the equation~\leqn{eq:orbifold}, with $\lambda_{1+} = \frac{1}{2} \left( 
1 + \gamma_5 \right) \lambda_1$ transforming like $A_{\mu}$ while $\sigma$ 
and $\lambda_{2+} = \frac{1}{2} \left( 1 + \gamma_5 \right) \lambda_2$ 
transform like $A_5$. The zero modes of $A_{\mu}$ and $\lambda_{1+}$ form a 
4D $N = 1$ gauge multiplet, while the other fields have no zero modes.

The remaining freedom concerns the location of the matter 
fields, the mechanism of breaking $SU(2)_R\times SU(2)_1\times SU(2)_2 
\rightarrow T_{3R}$, and the structure of the Higgs sector.

\subsection{Matter in the Bulk}

\begin{figure}
\begin{center} 
\begin{picture}(450,130)(-260,-85)
  \Line(-40,65)(20,15)
  \Line(20,15)(-40,-35)
  \Line(-40,-35)(-100,15)
  \Line(-100,15)(-40,65)
  \BCirc(-40, 65){15}
  \BCirc(20,15){15}
  \BCirc(-40,-35){15}
  \BCirc(-100,15){15}
  \Text(-40,65)[]{{\bf R}}
  \Text(20,15)[]{{\bf 1}}
  \Text(-100,15)[]{{\bf L}}
  \Text(-40, -35)[]{{\bf 2}}
  \Text(40,17)[lb]{$\tilde{\Psi}_L, \tilde{\Psi}'_L$}
  \Text(40,13)[lt]{$P_1, \bar{P}_1$}
  \Text(-120,17)[rb]{$\Psi_L, \Psi'_L$}
  \Text(-120,13)[rt]{$P_L, \bar{P}_L$}
  \Text(-8,42)[lb]{$\Phi_{R1},\bar{\Phi}_{R1}$}
  \Text(-8,-12)[lt]{$\Phi_{12},\bar{\Phi}_{12}$}
  \Text(-72,-12)[rt]{$\Phi_{2L},\bar{\Phi}_{2L}$}
  \Text(-72,42)[rb]{$\Phi_{LR},\bar{\Phi}_{LR}$}
  \Text(-40,-52)[ct]{$\tilde{\Psi}_R, \tilde{\Psi}'_R$}
  \Text(-40,-69)[ct]{$P_2, \bar{P}_2$}
  \Text(-40,84)[cb]{$P_R, \bar{P}_R$}
  \Text(-40,101)[cb]{$\Psi_R, \Psi'_R$}
\end{picture}
\caption{Quantum numbers of the matter and Higgs superfields under the 
four $SU(2)$ gauge groups in the model with matter in the bulk.}
\label{fig:transf}
\end{center}
\end{figure}
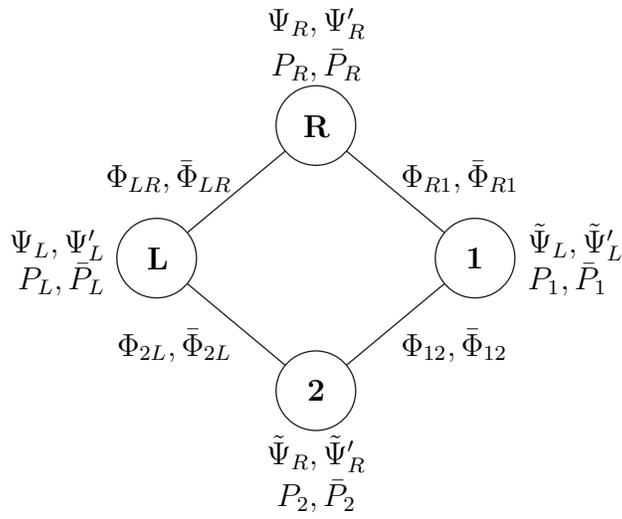

In our first model, the matter fields of the Standard Model arise as 
zero-modes of 5D fermions. One SM generation requires four 5D hypermultiplets, 
$(\Psi_L, \Psi'_L) \in (\4pl, \2pl, 1, 1, 1)$ and $(\Psi_R, \Psi'_R) \in 
(\bar{\4pl}, 1, \2pl, 1, 1)$, where in the brackets we list transformation 
properties under $SU(4) \times SU(2)_L \times SU(2)_R \times SU(2)_1 \times 
SU(2)_2$. Each 5D hypermultiplet $\Psi$ consists of a Dirac fermion 
$\psi$ and two complex scalars $\phi$ and $\phi^c$. The boundary conditions 
on the scalars are,
\beqa 
\phi (x^\mu, x_5) &=& +\phi (x^\mu, -x_5) = {\cal C}\,Z\,\phi (x^\mu, x_5+
2\pi R),\CR 
\phi^c (x^\mu, x_5) &=& -\phi^c (x^\mu, -x_5) = {\cal C}\,Z\,\phi^c (x^\mu, 
x_5+2\pi R), 
\eeqa{eq:orbiferm}
where $Z=~$diag$(1,1,1,-1)$, and ${\cal C}=\pm 1$ is the parity of the field 
$\Psi$. The fields $\psi_{+} = \frac{1}{2}(1 + \gamma_5)\psi$ and $\psi_{-} = 
\frac{1}{2} (1 - \gamma_5)\psi$ have the same boundary conditions as $\phi$ 
and $\phi^c$, respectively. The fields $\phi^c$ and $\psi_{-}$ have no 4D zero 
modes. The fields $\phi$ and $\psi_{+}$ each have a zero mode which together 
form a 4D $N=1$ chiral multiplet. The gauge charges of the zero modes of 
$\Psi$ under $SU(3)\times U(1)_{B-L}$ depend on its parity: for ${\cal C}=+1$ 
the zero modes transform as a quark or an antiquark, while for ${\cal C}=-1$ 
they transform as a lepton or an antilepton. We assign ${\cal C}=+1$ to the 
fields $\Psi_L$ and $\Psi_R$ and ${\cal C}=-1$ to the fields $\Psi'_L$ and 
$\Psi'_R$. As we already discussed in section \ref{bc}, to be able to impose 
a discrete symmetry necessary for a correct prediction of the weak mixing 
angle we need to introduce four more 5D fields per SM generation: 
$(\tilde{\Psi}_L, \tilde{\Psi}'_L) \in (\4pl, 1, 1, \2pl, 1)$ and 
$(\tilde{\Psi}_R, \tilde{\Psi}'_R) \in (\4pl, 1, 1, 1, \2pl)$. The zero modes 
of these fields form three additional ``spectator generations'', which however 
acquire masses at the scale $M_c'$ as we will show below.

The Higgs sector of the model consists of:
\begin{itemize} 
\item Eight 4D chiral superfields\footnote{It is easy to construct 
phenomenologically viable models with fewer 4D bidoublet Higgses. We choose
the structure presented here because of its pleasing symmetry, see
Fig.~\ref{fig:transf}.}, $\Phi_{ij}$ and $\bar{\Phi}_{ij}$, where
$(ij)= LR, R1, 12, 2L$. These fields are localized on the 3-1 boundary.
They transform as bidoublets (or ``link fields'')
under the corresponding $SU(2)$ groups, and are singlets of $SU(4)$.    
\item Eight 5D hypermultiplets, $P_i$ and $\bar{P}_i$, $i=1,2,L,R$. These
fields transform as $\4pl$ and $\bar{\4pl}$, respectively, under $SU(4)$, 
and as doublets under the corresponding $SU(2)$'s. The boundary conditions 
imposed on these fields are identical to~\leqn{eq:orbiferm} with 
${\cal C}=-1$. For example, the zero modes of the $P_R$ and $\bar{P}_R$
transform like 4D $N=1$ chiral multiplets that have the quantum numbers of a 
right-handed lepton doublet and a right-handed antilepton doublet,
respectively.
\end{itemize}

The matter and Higgs superfields of the model and their $SU(2)$ quantum 
numbers are conveniently summarized in Fig.~\ref{fig:transf}. 

We assume that the scalar components of $\Phi_{1R}$, $\bar{\Phi}_{1R}$, 
$\Phi_{12}$ and $\bar{\Phi}_{12}$ acquire diagonal vacuum expectation 
values (vevs) at the scale $M_c'$, breaking $SU(2)_R\times SU(2)_1 \times 
SU(2)_2$ down to the diagonal $SU(2)$ subgroup. (We do {\it not} break any of
the $SU(2)$'s by boundary conditions in this model.) The vev of the scalar 
component of $P_R$ at the same scale breaks the product of this $SU(2)$ and
$U(1)_{B-L}$ down to the SM hypercharge group $U(1)_Y$. Finally, the SM
electroweak symmetry breaking is achieved by the vevs of $\phi_u\equiv
\Phi_{LR}$ and $\phi_d\equiv\bar{\Phi}_{LR}$, for which we assume the 
pattern\footnote{This pattern of the electroweak symmetry breaking is chosen
to enable us to impose a discrete symmetry protecting vanishing neutrino 
Yukawa couplings, see below.}
\beq
\phi_u \,=\, \left( \begin{array}{cc} 
  v_u & 0 \\ 
    0 & 0 
       \end{array} 
\right), \hskip2cm
\phi_d \,=\, \left( \begin{array}{cc} 
    0 & 0 \\ 
    0 & v_d 
       \end{array} 
\right).
\eeq{vevs} 
The rest of the Higgs fields do not acquire vevs.

Let us analyze the pattern of masses for the matter fields. To simplify the 
analysis, we assume that the Yukawa couplings satisfy a discrete 
$Z_2$ symmetry under which the ``spectator generation'' superfields
($\tilde{\Psi}_{L,R}$ and $\tilde{\Psi}'_{L,R}$) have a charge $-1$ and the 
rest of the fields are invariant. Then, the zero modes of the ``spectator 
generations'' get masses 
at the scale $M_c'$ through the superpotential couplings $\tilde{\Psi}_L 
\Phi_{12} \tilde{\Psi}_R+\tilde{\Psi}'_L \Phi_{12}\tilde{\Psi}'_R$, while the 
three ordinary generations get masses at the weak scale through their couplings
to $\phi_u$ and $\phi_d$. In addition, the right-handed neutrino gets a 
Majorana mass of order 10 GeV through a non-renormalizable coupling $(\nu^c 
P_R)^2/M_s$.  
     
The main challenge in the flavor sector of the model is to explain the
smallness of neutrino Dirac masses: successful phenomenology requires
$m_\nu^D\approx 10^4$ eV. We take the view that the neutrino Yukawa couplings
vanish exactly due to a discrete symmetry. We postulate a symmetry 
$\phi_d \goesto e^{i\,\pi/2} \phi_d$, $\Psi'_L \goesto e^{i\,3\pi/2}$, with 
all the other fields being invariant. This symmetry allows 
superpotential Yukawa couplings $L\phi_d e^c + Q\phi_u u^c$, but forbids 
$L\phi_u\nu^c$ and $Q\phi_d d^c$. The down-type quarks acquire their masses 
as a result of spontaneous supersymmetry breaking on the 3-1 boundary at a 
scale of order $M_c$. The Kahler potential term $\int d^4\theta X^\dagger 
Q\phi_u^\dagger d^c/M_s$, where $X$ is the supersymmetry breaking spurion, 
induces a 
down quark Yukawa coupling suppressed by $M_c/M_s \sim 0.01$ -- the correct 
order of magnitude to explain the hierarchy between the top and bottom masses.
The corresponding term for the neutrino, $\int d^4\theta X^\dagger L
\phi_d^\dagger \nu^c/M_s$, is still forbidden by the discrete symmetry. A 
neutrino Dirac mass of the right order of magnitude can be generated by  
known higher-dimensional mechanisms~\cite{neutrinos}.

Apart from the boundary-localized Yukawa couplings, the model possesses a 
cyclic symmetry interchanging the four $SU(2)$ factors, implying $b_R'=3b_L'$.
Using the results of Section~\ref{loops}, we obtain
\beq
\sin^2 \theta (M_Z) = 0.231 - 0.0031 \ln \left(\frac{M_c'}{{\rm 200~GeV}}
\right) \pm 0.005,
\eeq{model1}
where we have assumed a 2\% uncertainty on the prediction 
as explained at the end of Section~\ref{loops}. Our model contains the exotic 
gauge bosons $W_{LR}'$ and $Z'_{LR}$ with masses of order $M_c'$; experimental 
limits on such bosons in the canonical left-right symmetric model~\cite{pdg} 
require that their mass be higher than about 800 GeV. In our model, the 
limits are somewhat higher due to a stronger coupling of the $B-L$ component
of $Z'_{LR}$. However, the uncertainty in the prediction~\leqn{model1} is 
sufficiently high to allow values of $M_c'$ as high as 5 TeV at $2\sigma$
level, so the model
is phenomenologically consistent. On the other hand, the 
non-supersymmetric version of the same model does not look viable, since
it requires $M_c'\lsim 650$ GeV to reproduce the 
experimentally measured value of the weak mixing angle within $2\sigma$.  

There are new sources of flavor-changing neutral currents in our model.
The tree-level exchanges of the additional neutral Higgs bosons in $\phi_u$
and $\phi_d$ contribute to $K\bar{K}$ and $D\bar{D}$ mass differences. 
Suppressing this contribution requires that the mixing between the third and 
the first two generations in the right-handed sector be about as small as
in the left-handed sector. In additon, there is a contribution from the
box diagrams involving $W_R$ gauge bosons. However, the small mixings of
the third generation and the large masses of the $W_R$ bosons ensure that 
this effect is harmless. 

\subsection{Matter on the Boundary}

We now consider the possibility that both the breaking of $SU(4)$ to 
$SU(3) \times U(1)_{B-L}$ and the breaking of $SU(2)_R \times SU(2)_1 \times
SU(2)_2$ to $U(1)_R \times U(1)_1 \times U(1)_2$ are accomplished using 
boundary conditions\footnote{A 5D left-right symmetric model with $SU(2)_R$ 
broken by boundary conditions was also considered in~\cite{Mohapatra}.}. 
For each $SU(2)$ field the appropriate pattern of breaking is obtained by 
demanding 
\beqa 
A^{\mu}_i (x^\mu, x_5) &=& +A^{\mu}_i (x^\mu, -x_5) = T_i\,A^{\mu}_i 
(x^\mu, x_5+2\pi R)\,T_i^{-1},\CR 
A^{5}_i (x^\mu, x_5) &=& -A^5_i (x^\mu, -x_5) = T_i\,A^5_i 
(x^\mu, x_5+2\pi R)\,T_i^{-1},
\eeqa{eq:orbifold2}                                                            
where $\mu = 1 \ldots 4$ and the index $i$ denotes the $SU(2)$ gauge group 
and runs over $L,R,1$ and 2. We choose $T_L =$ diag$(+,+)$ for $SU(2)_L$ 
reflecting the fact that it is not broken by boundary conditions; for the
other $SU(2)$ groups, we choose $T_R = T_1 = T_2 =$ diag$(+,-)$. The quark 
and lepton fields are localized on the boundary where both the $SU(4)$ and the
$SU(2)^3$ symmetries are broken. The cyclic exchange symmetry between the 
$SU(2)$ groups is also broken at this point. We assume that the quark and 
lepton fields transform as three generations of $\Psi_L\in (\4pl,\2pl,1,1,1)$ 
and $\Psi_R\in(\bar{\4pl},1,\2pl^{*},1,1)$, where for notational simplicity we 
show the transformation properties under $SU(4) \times SU(2)_L \times SU(2)_R 
\times SU(2)_1 \times SU(2)_2$. (The transformation properties under the group
that remains unbroken on the boundary, $[SU(3) \times U(1)_{B-L}] \times 
SU(2)_L \times U(1)_R \times U(1)_1 \times U(1)_2$, can be easily obtained 
from these.) No ``spectator'' generations are required in this model.

The model we consider is supersymmetric. The 
additional fields of the higher dimensional gauge multiplet transform as in 
the equation above, with $\lambda_{1+}$ transforming like $A_{\mu}$ and 
$\lambda_{2+}$ and $\sigma$ transforming like $A_5$. As before the fields 
$A_{\mu}$ and $\lambda_{1+}$ have zero modes which combine to form a 4D
$N=1$ gauge multiplet.  The matter fields on the boundary become 4D $N=1$ 
chiral superfields.

We now consider the Higgs sector of the theory. The Higgs fields are assumed 
to live in the bulk of the space. We wish to break $U(1)_R \times U(1)_1 
\times U(1)_2$ to the diagonal $U(1)_R$. We therefore introduce pairs of Higgs 
hypermultiplets in the bulk which we denote by $\Phi_{ij}$ and 
$\bar{\Phi}_{ij}$, where $(ij)= LR, R1, 12, 2L$. These behave like link fields,
transforming as bidoublets under $SU(2)_i \times SU(2)_{j}$ but as singlets 
under all other gauge groups. For example ${\Phi}_{LR}$ and $\bar{\Phi}_{LR}$ 
transform as $({\2pl}^*,\2pl)$ and $(\2pl,{\2pl}^*)$ respectively under
$SU(2)_L \times SU(2)_R$, but as singlets under the remaining gauge groups. 
Each hypermultiplet {\bf $\Phi_{ij}$} consists of a Dirac fermion $\psi_{ij}$ 
and two complex scalars $\phi_{ij}$ and ${\phi}^c_{ij}$. The transformation
properties of the scalars under the orbifold are given by
\beqa 
\phi_{ij} (x^\mu, x_5) &=& +\phi_{ij} (x^\mu, -x_5) = T_i\, \phi_{ij} (x^\mu,
x_5+2\pi R)\, T_j^{-1}, 
\CR {\phi}^c_{ij} (x^\mu, x_5) &=& - {\phi}^c_{ij} (x^\mu, -x_5) = 
T_i\, {\phi}^c_{ij}
(x^\mu, x_5+2\pi R)\, T_j^{-1}.
\eeqa{eq:orbifold3}   
The fermions $\psi_{ij,+}$ transform exactly like the $\phi_{ij}$, while the 
fermions $\psi_{ij,-}$ transform like ${\phi}^c_{ij}$. The fields 
${\phi}_{ij}$ and $\psi_{ij,+}$ have zero modes which combine to form 4D 
$N=1$ chiral multiplets. The other fields have no zero modes. Similarly each 
hypermultiplet $\bar{\Phi}_{ij}$ consists of a Dirac fermion $\bar{\psi}_{ij}$ 
and two complex scalars $\bar{\phi}_{ij}$ and ${\bar{\phi}}^c_{ij}$. While 
$\bar{\phi}_{ij}$ and $\bar{\psi}_{ij,+}$ transform exactly like the 
$\phi_{ij}$ under the orbifold and have zero modes, the remaining fields 
transform like ${\phi}^c_{ij}$ and have no zero modes. The zero modes of 
$\bar{\Phi}_{ij}$ form a 4D $N=1$ chiral multiplet that is vector-like
with respect to the zero modes of ${\Phi}_{ij}$.

It is possible to write superpotential terms for the even components of the 
$\Phi$ and $\bar{\Phi}$ fields on the symmetry breaking boundary. These can 
be used to generate a potential for the even components of [$\Phi_{R1}$, 
$\bar{\Phi}_{R1}$] and [$\Phi_{12}$, $\bar{\Phi}_{12}$] that breaks $U(1)_R
\times U(1)_1 \times U(1)_2$ down to the diagonal $U(1)_R$. The even 
components of $\Phi_{LR}$ and $\bar{\Phi}_{LR}$ correspond to the up type 
Higgs and down type Higgs chiral multiplets of the Minimal Supersymmetric
Standard Model (MSSM).

There remains the breaking of $U(1)_R \times U(1)_{B-L}$ down to $U(1)_Y$. 
For this purpose we introduce pairs of hypermultiplets in the bulk which we 
denote by $P_i$ and $\bar{P}_i$. $P_i$ and $\bar{P}_i$ transform as $\4pl$ 
and $\bar{\4pl}$ under $SU(4)$ respectively, as doublets under $SU(2)_i$ and 
as singlets under the other $SU(2)$ groups. Each $P_i$ consists of two complex 
scalars $p_i$ and $p^c_i$, and a Dirac fermion $\psi_{Pi}$. The 
transformation properties of the scalars under the orbifold are,
\beqa 
p_{i} (x^\mu, x_5) &=& +p_{i} (x^\mu, -x_5) = -Z T_i\, p_{i} (x^\mu,
x_5+2\pi R), 
\CR p^c_{i} (x^\mu, x_5) &=& - p^c_i (x^\mu, -x_5) 
= -Z T_i \, p^c_{i} (x^\mu, x_5+2\pi R). 
\eeqa{eq:orbifold4} 
The fermions $\psi_{Pi,+}$ transform like $p_i$ while the fermions 
$\psi_{Pi,-}$ transform like $p^c_i$. While $\psi_{Pi,+}$ and $p_i$ have zero 
modes which combine to form 4D $N=1$ chiral multiplets, the other fields have 
no zero modes. Similarly each of the hypermultiplets $\bar{P}_i$ consists of 
two complex scalars $\bar{p}_i$ and $\bar{p}^c_i$, and a Dirac fermion 
$\bar{\psi}_{Pi}$. While $\bar{p}_i$ and $\bar{\psi}_{Pi,+}$ share the same 
orbifold transformation properties as $p_i$ and have zero modes which combine 
to 
form 4D $N=1$ chiral multiplets, the other fields have no zero modes. The zero
modes of $P_i$ have the gauge quantum numbers of a neutrino of $SU(2)_i$
and a down quark of $SU(2)_i$ and $SU(3)$. The zero modes of $\bar{P}_i$ are 
vectorlike with respect to those of $P_i$. Then it is possible to write a 
potential for the zero modes of $P_R$ and $\bar{P}_R$ (which have the quantum 
numbers of a right handed neutrino and a right handed anti-neutrino) on the 
boundary which breaks $U(1)_R \times U(1)_{B-L}$ down to $U(1)_Y$.
\begin{figure}
\begin{center} 
\begin{picture}(450,130)(-260,-74)
  \Line(-40,65)(20,15)
  \Line(20,15)(-40,-35)
  \Line(-40,-35)(-100,15)
  \Line(-100,15)(-40,65)
  \BCirc(-40, 65){15}
  \BCirc(20,15){15}
  \BCirc(-40,-35){15}
  \BCirc(-100,15){15}
  \Text(-40,65)[]{{\bf R}}
  \Text(20,15)[]{{\bf 1}}
  \Text(-100,15)[]{{\bf L}}
  \Text(-40, -35)[]{{\bf 2}}
  \Text(40,15)[lc]{$P_1, \bar{P}_1$}
  \Text(-130,18)[rb]{$\Psi_L$}
  \Text(-120,13)[rt]{$P_L, \bar{P}_L$}
  \Text(-8,42)[lb]{$\Phi_{R1},\bar{\Phi}_{R1}$}
  \Text(-8,-12)[lt]{$\Phi_{12},\bar{\Phi}_{12}$}
  \Text(-72,-12)[rt]{$\Phi_{2L},\bar{\Phi}_{2L}$}
  \Text(-72,42)[rb]{$\Phi_{LR},\bar{\Phi}_{LR}$}
  \Text(-40,-54)[ct]{$P_2, \bar{P}_2$}
  \Text(-40,84)[cb]{$P_R, \bar{P}_R$}
  \Text(-40,101)[cb]{$\Psi_R$}
\end{picture}
\caption{Quantum numbers of the matter and Higgs superfields under the 
four $SU(2)$ gauge groups in the model with matter on the boundary.}
\label{fig:2}
\end{center}
\end{figure}
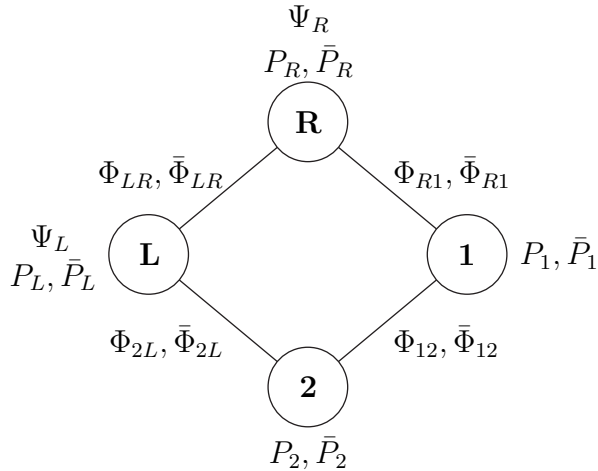

The Higgs and matter fields of the model and their $SU(2)$ quantum numbers
are summarized in Figure~\ref{fig:2}. It is straightforward to generate
fermion masses from 
Yukawa couplings between the Higgs fields and the matter fields on the 
symmetry breaking boundary. 
It is also not difficult to find a discrete symmetry 
which adequately suppresses the Dirac
mass of the neutrino since there are no longer
any mass relations from the larger gauge symmetry in the bulk.

From Eq.~\leqn{eq:s^2master} we can obtain an expression for the 
compactification scale $M_c$. For concreteness we assume that supersymmetry 
is broken at the scale $M_c'$ and therefore use the Standard Model expressions 
for $b_Y, b_2$ and $b_3$ when running below the compactification scale. Using 
the fact that bulk hypermultiplets do not contribute to any of the $b'$s we 
find that $b'_R - 3 b_2' = -12$. Then the expression for the weak mixing angle 
becomes
\beq 
\sin^2 \theta (M_Z) = 0.231 - 0.0065 \ln \left(\frac{M_c'}{{\rm 1.74~TeV}} 
\right) \pm 0.005.
\eeq{model2}    
Even without invoking the uncertainties, the values of $M_c'$ required to 
reproduce the experimentally measured value of $\sin^2\theta$ are 
sufficiently high to evade experimental constraints. The central value of the 
compactification scale is given by $M_c = \pi M_c' = 5.5$ TeV.

\section{Conclusions} \label{conclusions}

The first theory to unify quarks with leptons was based on the group
$SU(4) \times SU(2)_L \times SU(2)_R$, and provided an elegant understanding 
of the fermion gauge quantum numbers \cite{PS}. We have shown that an 
extension to $SU(4) \times SU(2)^4$, considered in the past by Hung, Buras 
and Bjorken~\cite{petite}, allows this Pati-Salam structure to be realized at 
the TeV scale. An appropriate embedding of the Standard Model generators 
leads to a tree-level prediction for the weak mixing 
angle, $\sin^2 \theta = 0.239$, which is remarkably close to data. Such a
TeV scale unification can be combined with extra-dimensional
solutions of the hierarchy problem, by adding very large
dimensions \cite{ADD}, or by adding an extra dimension with a warp
factor \cite{RS} and placing the extended Pati-Salam sector on the TeV
brane. 

We do not view the extension from $SU(2)^2$ to $SU(2)^4$ as a major
complication of the theory. In both cases a discrete symmetry between the 
$SU(2)$ factors is necessary to obtain a prediction for the weak mixing 
angle, and this symmetry must be broken, protecting only $SU(2)_L$ to lower 
energies. Various interpretations of the four $SU(2)$ factors are possible. 
For example, in the model of section 4.1 the Pati-Salam structure is simply
repeated:  $SU(2)_L \times SU(2)_R \times SU(2)'_L \times SU(2)'_R$.
Another possibility, not pursued in this paper, is to have a different 
$SU(2)_{R_a}$ for each generation $a$: $SU(2)_L \times 
\prod_a SU(2)_{R_a}$. The usual $SU(2)_R$ is then just the diagonal sum
of the $SU(2)_{R_a}$, leading
to the desired prediction for the weak mixing angle.

We have realized the $SU(4) \times SU(2)^4$ symmetry in 5 dimensions, with 
boundary
conditions in the compact fifth dimension breaking 
$SU(4) \rightarrow SU(3) \times U(1)_{B-L}$. This facilitates
imposing the discrete symmetry among the $SU(2)$ factors,
removes the powerful constraint from $K_L \rightarrow \mu e$ on the
masses of the charged $SU(4)$ gauge bosons, leads to Yukawa
couplings which are not $SU(4)$ symmetric and allows baryon number to be 
a symmetry of the theory. Providing the five
dimensional effective theory is valid up to a scale where the gauge
couplings approach strong coupling, the prediction for the weak mixing
angle is under control \cite{HN5,HN3}, and our result, including
radiative corrections and uncertainties, is shown in~\leqn{s^2num} 
and~\leqn{s^2numSUSY} for this class of theories.

There are many possibilities for the breaking pattern of $SU(2)^3
\times U(1)_{B-L} \rightarrow U(1)_Y$, leading to
model dependence in the weak mixing angle prediction and in the signatures 
at
future collider experiments. We have given two explicit supersymmetric
models; one having the three $SU(2)$s broken by boundary conditions,
and the other having the symmetry breaking entirely from the Higgs
mechanism. In both cases, the weak mixing angle is successfully
predicted, although the central values for the compactification scale
differ: 5.5 TeV for boundary condition breaking, and 600 GeV for Higgs
breaking. However, there is an order of magnitude uncertainty in these
values of the compactification scale arising from the uncertainty 
in the prediction of the weak mixing angle. While theories of $SU(4)
\times SU(2)^4$ unification predict many new phenomena at the TeV
scale, including heavy $W'$ and $Z'$ bosons and KK modes for all gauge
bosons, there is a significant uncertainty in the precise energy
threshold for this new physics. The reach of the Tevatron and the LHC for
the neutral $Z'$ gauge boson will be high: its $B-L$ component couples with 
a QCD strength gauge coupling, it is singly produced in $q\bar{q}$ collisions 
(or pair-produced in $gg$ collisions), and decays readily to charged quark or
lepton pairs. The quark and lepton branching ratios will reveal that this 
$Z'$ boson is coupled to $\frac{1}{2}(B-L) - T_{3R}$, while its production 
cross section will indicate that $SU(4)$ unification occurs at a low scale.

\section*{Acknowledgments} 
We would like to thank Kaustubh Agashe and Christophe Grojean for helpful
comments. The authors are supported by the Director, Office 
of Science, Office of High Energy and Nuclear Physics, of the U. S. Department 
of Energy under Contract DE-AC03-76SF00098, and by the National Science
Foundation under grant PHY-00-98840.

\end{document}